\documentclass[12pt]{article}

\usepackage{scicite}

\usepackage{times}
\usepackage{braket}
\usepackage{bm}
\usepackage{amsmath}
\usepackage{graphicx}
\graphicspath{{./figures/}} % this sets the actual path of figures
\usepackage{upgreek}    % Allow non-italic greek letters       
\usepackage{mhchem}

\newenvironment{sciabstract}{%
\begin{quote} \bf}
{\end{quote}}

\newcounter{lastnote}

\title{Optically Addressable Molecular Spins at 2D Surfaces}
\author
{Xuankai Zhou,$^{1\ast}$ Yan-Tung Kong,$^{1\ast}$  Cheuk Kit Cheung,$^{1\ast}$ Guodong Bian,$^{2,4}$\\ Reda Moukaouine,$^{2,5}$ King Cho Wong,$^{1}$ Yumeng Sun,$^{1}$ Cheng-I Ho,$^{1}$ \\Vladislav Bushmakin,$^{1}$ Nils Gross,$^{3}$ Chun-Chieh Yen,$^{3}$ Tim Priessnitz,$^{3}$ \\Malik Lenger,$^{1}$ Sreehari Jayaram, $^{1}$ Takashi Taniguchi,$^{8}$ Kenji Watanabe,$^{9}$\\ Anton Pershin,$^{2,6}$ Ruoming Peng,$^{1\dagger}$ Ádám Gali,$^{2,6,7}$  Jurgen Smet,$^{3}$\\ Jörg Wrachtrup$^{1,3}$\\
\\
\normalsize{$^{1}$3. Physikalisches Institut, University of Stuttgart, 70569 Stuttgart, Germany}
\\
\normalsize{$^{2}$HUN-REN Wigner Research Centre for Physics, Institute for Solid State}\\
\normalsize{Physics and Optics, P.O. Box 49, H-1525, Budapest, Hungary}
\\
\normalsize{$^{3}$Max Planck Institute for Solid State Research, 70569 Stuttgart, Germany}
\\
\normalsize{$^{4}$School of Chemistry, University of Birmingham, B15 2TT, Edgbaston, Birmingham, UK}
\\
\normalsize{$^{5}$György Hevesy Doctoral School, ELTE Eötvös Loránd University, Institute of Chemistry,}\\
\normalsize{H-1117 Budapest, Hungary}
\\
\normalsize{$^{6}$Department of Atomic Physics, Institute of Physics, Budapest University of Technology}\\
\normalsize{and Economics, M\H{u}egyetem rakpart 3, 1111 Budapest, Hungary}
\\
\normalsize{$^{7}$MTA-WFK Lendület “Momentum” Semiconductor Nanostructures Research Group,}\\
\normalsize{P.O. Box 49, H-1525, Budapest, Hungary}
\\
\normalsize{$^{8}$ Research Center for Materials Nanoarchitectonics, National Institute for Materials}\\
\normalsize{Science, 1-1 Namiki, Tsukuba 305-0044, Japan}
\\
\normalsize{$^{9}$ Research Center for Electronic and Optical Materials, National Institute for Materials}\\
\normalsize{Science, 1-1 Namiki, Tsukuba 305-0044, Japan}
\\
\\
\\
\normalsize{$^\ast$Equal contributions}
\\
\normalsize{$^\dagger$E-mail: ruoming.peng@pi3.uni-stuttgart.de}
}
\date{}

\begin{document} 


\baselineskip24pt


\maketitle

\pagebreak
\begin{sciabstract}
Optically addressable spins at material surfaces have represented a long-standing ambition in quantum sensing, providing atomic resolution and quantum-limited sensitivity. However, they are constrained by a finite depth at which the quantum spins can be stabilized. Here, we demonstrate a hybrid molecular-2D architecture that realizes quantum spin sensors directly on top of the surface. By anchoring spin-active molecules onto hexagonal boron nitride (hBN), we eliminate the depth of the quantum sensor while also exhibiting robust spin properties from 4 K to room temperature (RT). The Hahn-echo spin coherence time exceeds $T_2 = 3.4~\upmu$s at 4~K, outperforming values in bulk organic crystals and overturning the prevailing expectation that spin inevitably deteriorates upon approaching the surface. By chemically tuning the molecule through deuteration, $T_2$ improves by more than 10-fold, and under dynamic decoupling, coherence is prolonged to the intrinsic lifetime limit, exceeding 300~$\upmu$s. Proximal proton spins and the magnetic response of two-dimensional magnets beneath the hBN layer have been detected at RT. These molecular spins form surface quantum sensors with long coherence, optical addressability, and interfacial versatility, enabling a scalable, adaptable architecture beyond what conventional solid-state platforms offer.

\end{sciabstract}

% In setting up this template for *Science* papers, we've used both
% the \section* command and the \paragraph* command for topical
% divisions.  Which you use will of course depend on the type of paper
% you're writing.  Review Articles tend to have displayed headings, for
% which \section* is more appropriate; Research Articles, when they have
% formal topical divisions at all, tend to signal them with bold text
% that runs into the paragraph, for which \paragraph* is the right
% choice.  Either way, use the asterisk (*) modifier, as shown, to

\section*{Introduction}
Quantum sensing has emerged as a powerful technology for probing matter at the nanoscale \cite{Degen2017}. Applications range from single nuclear magnetic resonance detection to the direct imaging of emergent phases in quantum materials, such as magnetism and superconductivity \cite{Du2024,Casola2018}. By combining long-lived quantum coherence with nanometer spatial resolution, spin-based platforms provide unique opportunities to access the electromagnetic responses at the atomic scale that remain out of reach for conventional probes \cite{Balasubramanian2008,Maletinsky2012,Staudacher2013,Thiel2019,Bian2021,Song2021,Palm2024}. Advancing the capabilities of quantum sensing requires that spin sensors are positioned ever closer to the target, ideally directly at the surface itself \cite{Schirhagl2014}. However, this longstanding pursuit has faced persistent difficulties: defects, adsorbates, and surface terminations introduce a variety of noise sources that significantly degrade the spin coherence, charge stability, and sensitivity, in particular when the spin-to-surface distance drops below 10~nm \cite{Romach2015,Bluvstein2019,Sangtawesin2019}. Despite recent advances in bulk solid-state hosts like diamond and SiC \cite{Rodgers2021,Wolfowicz2021,Li2025}, the fundamental trade-off between depth and spin properties remains, motivating the exploration of alternative platforms.

Two-dimensional (2D) layered materials offer intrinsic structural stability and atomically clean interfaces with minimal surface-induced charge traps and no dangling bonds \cite{Dean2010} and are therefore considered promising candidates for quantum sensing at surfaces \cite{Tran2015,Vaidya2023,Fang2024}. Multiple optically addressable spin defects have been identified in 2D hexagonal Boron Nitride (hBN) \cite{Gottscholl2020,Mendelson2020,Chejanovsky2021,Stern2022,Gao2025}. However, their spin properties are limited by the dense nuclear spin environment of hBN \cite{Haykal2022} and the defects are often either too dim for efficient optical readout or challenging to reproducibly create due to the unclear defect structures \cite{Vaidya2023,Liu2022}. It also remains a subject under debate whether such spin defects can be stabilized at the hBN surface, i.e. in the top hBN layers \cite{Dai2023}. So far, it is observed that the spin properties of the defects in hBN also degrade significantly as they approach the surface to less than 10~nm \cite{Durand2023}. 

In parallel, molecular spin systems have also been extensively explored for a variety of quantum applications \cite{Wrachtrup1993,Bayliss2020,Serrano2022,Mena2024,Weiss2025}. While these systems offer chemically tunable quantum properties \cite{Mann2025}, they typically suffer from limited stability. They must be dispersed in bulk organic host crystals, where their spin properties again start to degrade when thinned down to below 100~nm \cite{Mena2024}. This renders surface integration and nanoscale positioning impractical. Moreover, many of these molecular spin candidates are only accessible at cryogenic temperatures with resonant excitation \cite{Bayliss2020,Serrano2022,Weiss2025}. That greatly limits their potential for robust quantum applications.

In order to overcome these limitations and realize stable spin systems with well-preserved spin-optical properties at the surface, we combine aromatic spin molecules-more specifically, pentacene (Pc)-with the 2D material hBN to construct a hybrid surface spin platform as illustrated in Fig.1a. Although such molecules have been deposited on hBN for high-quality organic electronics \cite{Kang2014,Zhang2016,Guender2020,Amsterdam2020}, their potential for quantum information processing at surfaces has remained entirely unexplored. The hBN can serve multiple purposes for molecular quantum sensing: (1) It provides a chemically inert and structurally stable substrate that protects the spin molecules from environmental degradation; (2) Its wide bandgap suppresses charge transfer from the Pc molecules, thereby maintaining the spin-triplet state; and (3) The lattice of B–N bonds in hBN closely matches the atomic spacing of C–C bonds in aromatic molecules, promoting the self-assembly of molecular arrays on the hBN surface; (4) Finally, it hosts a variety of surface defects that interact with and couple to Pc. This modifies the Pc energy levels of coupled Pc and mitigates unwanted annihilation processes from surrounding molecules. Together, these properties make hBN an ideal platform for integrating aromatic molecular spins toward the realization of quantum sensing at surfaces with prolonged spin coherence and enhanced stability.

\section*{Optically Addressable Surface Spins}

Pc–hBN devices are prepared via standard drop-casting/dip-coating techniques (see Supplementary Section 2). Pc is first dissolved in an organic solvent (e.g., chlorobenzene or toluene) with a concentration of $\sim$10 mg/L, and then deposited onto the hBN flake positioned adjacent to a metal stripline. Because of the close match between the C-C and B-N bond lengths, as well as the hexagonal bond configuration of the molecule's aromatic rings and the hBN lattice, two molecular orientations are favored at the surface, as illustrated in Fig.1a: face-on and edge-on. In the face-on phase, densely packed molecules exhibit strong spin diffusion and triplet annihilation, which quench spin triplet states. In contrast, edge-on Pc, stabilized by native or intentionally introduced defects in the hBN surface layer, forms more decoupled molecules with weaker intermolecular coupling. As illustrated in Fig.1a, these edge-on molecules with a height of only 1.45~nm can act as coherent surface spins. Interactions with the hBN surface defects can distort the molecular orbitals, such as the highest occupied molecular orbitals (HOMO) of
the triplet spins (Fig.1b), and the photoluminescence (PL) spectrum shows features of isolated Pc molecules—an essential prerequisite for maintaining active triplet spin states \cite{Wrachtrup1993,Mena2024}. Specifically, we observe a zero-phonon line near 580 nm and a phonon sideband that matches the intrinsic isolated molecular vibrational mode (Fig. 1c). 

%We now turn our attention to the role of hBN surface conditions to stabilize the coherent surface spins. Samples with three different surface treatments were studied in order to identify a correlation between the surface defect density and the occurrence of spots with optically detected magnetic resonance (ODMR): pristine hBN with no treatment, electron irradiated hBN samples, and O$_2$-plasma treated samples. The hBN samples with heavy O$_2$-plasma bombardment show a strong increase in the number of ODMR spots, whereas in electron-irradiated samples, only a few spots are observed. In pristine high-quality hBN, almost none can be detected (See Supplementary Section 2 for more details). %By applying 20-minute O$_2$-plasma treatment, we successfully achieve full coverage of Pc molecules on hBN, as demonstrated in Fig.1c, with every spot in the confocal map exhibiting a clear ODMR signal. Motivated by this trend, we employed a hard mask to deterministically generate defective regions in hBN using ICP-RIE etching. As shown in Fig.1d, pentacene molecules selectively anchor to these patterned defect arrays, producing emission spots that are spatially aligned with the etched pattern. Although the Pc molecules in ICP-etched regions exhibit noticeably reduced photostability compared with those on plasma-treated hBN, likely due to unwanted damage introduced by the more aggressive etching process, the approach nonetheless demonstrates a viable route toward nanoscale deterministic placement of Pc spin molecules on hBN.

The left inset of Fig.1d summarizes the observed spin transitions and electronic structure of an isolated Pc molecule. It features a singlet ground state ($\mathrm{S_0}$) and a series of excited states, including a long-lived triplet manifold. Upon 532-nm laser excitation, the molecule is excited from the singlet ground state to an excited singlet state ($\mathrm{S_1}$). The excited population can subsequently relax back either radiatively to the ground state via fluorescence, or non-radiatively by intersystem crossing (ISC) to a metastable triplet state ($\mathrm{T}$) with high yield, as commonly observed in organic host matrices. Due to the strong zero-field splitting (ZFS) intrinsic to Pc molecules and the additional symmetry reduction induced by the attachment to the hBN surface, the triplet state degeneracy is lifted. This results in three distinct sublevels: $\ket{\mathrm{T_x}}$, $\ket{\mathrm{T_y}}$, and $\ket{\mathrm{T_z}}$.

Continuous-wave (cw) ODMR measurements performed under ambient conditions reveal clear resonance features when the microwave frequency matches the energy separation between the triplet sublevels (schematically illustrated in the right inset of Fig.1d). The Pc on plasma-treated or electron-irradiated hBN exhibits enhanced ZFS constants, with the $\ket{\mathrm{T_y}} \leftrightarrow \ket{\mathrm{T_z}}$ and $\ket{\mathrm{T_y}} \leftrightarrow \ket{\mathrm{T_x}}$ transitions observed at 1433~MHz and 917~MHz, respectively. Using a double-resonance technique, holding one microwave tone at 1433 MHz and sweeping the second, we uncover an additional resonance near 2350 MHz, confirming access to all triplet sublevels (Fig.1d). Importantly, we observe a linewidth of approximately 5~MHz for the $\ket{\mathrm{T_y}} \leftrightarrow \ket{\mathrm{T_z}}$ transition. This value approaches the proton-limited linewidth typically reported for Pc in organic hosts. The molecular spins also show remarkable photostability, enabling cw-ODMR measurements lasting for days under ambient conditions and no changes for months at 4 K, far exceeding the stability reported in typical thin-film molecular systems. As shown in Supplementary Section 3, the time trace of photon counts shows a fast decay time of~2 hours and a slow decay exceeding two days, whereas the cw-ODMR contrast remains visible over the whole measurement duration.

We also manipulated spin states of Pc on hBN by driving Rabi oscillations at the $\ket{\mathrm{T_y}}\leftrightarrow\ket{\mathrm{T_z}}$ transition. At room temperature, coherent control was achieved with 7\% maximum contrast (Supplementary Section 7). It is still limited by the background fluorescence from defective hBN and uncoupled molecules. Hahn-echo measurements delivered a $T_2$ of 2.4 $\upmu$s at room temperature and 3.4~$\upmu$s at 4~K (Fig.1e), exceeding the previously reported values of Pc in p-terphenyl \cite{Wrachtrup1993}. Unlike Pc in organic crystals, where $T_2$ collapses as thin-film thickness is below $\sim200$~nm \cite{Mena2024}, Pc on hBN exhibits exceptional spin coherence, surpassing bulk values. AFM maps reveal ultrathin Pc clusters with a minimum thickness of 3~nm (Supplementary Section 4), confirming stabilization of Pc spin states at the interface. These results have established hBN as an exceptional surface host for aromatic molecular spins, with $T_2$ values surpassing those of all known hBN spin defects by more than an order of magnitude across a broad range of temperatures.

\section*{Hybrid Molecule-2D Framework}
We now turn our attention to the role of hBN surface conditions to stabilize the coherent surface spins. Samples with three different surface treatments were studied in order to identify a correlation between the surface defect density and the occurrence of spots with optically detected magnetic resonance (ODMR): pristine hBN with no treatment, electron irradiated hBN samples, and O$_2$-plasma treated samples. The hBN samples with heavy O$_2$-plasma bombardment show a strong increase in the number of ODMR spots, whereas in electron-irradiated samples, only a few spots are observed. In pristine high-quality hBN, almost none can be detected (See Supplementary Section 2 for more details).

For Pc embedded in organic host matrices, zero-field axial and rhombic splitting parameters $D$ and $E$ typically take on values of about 1400~MHz and 50~MHz, respectively \cite{Mena2024,Singh2025}. For Pc on hBN, $D$ and $E$ are significantly larger. We obtain values of 1891~MHz and 459~MHz at room temperature. Moreover, the triplet population distribution differs markedly from the triplet population in an organic host, where the $\ket{\mathrm{T_x}}$ state is predominantly occupied \cite{Mena2024}. We argue that these differences reflect the altered local environment and highlight the importance of hBN and its defects in anchoring the Pc molecules. 
%Indeed, DFT calculations confirm that Pc molecules can couple to defects of the hBN surface and yield larger $D$ and $E$ values compared to free Pc molecules (see Supplementary section 9). In the calculations, Pc adopts the edge-on configuration for V$_B$-V$_N$ divacancies, as illustrated in Fig.2B. The spin density distribution is shifted toward the hBN plane, which further reduces the symmetry of the molecule. 
Indeed, DFT calculations show that Pc molecules can establish chemical bonds with defects on the hBN surface, leading to larger zero-field splitting parameters $D$ and $E$ than in free Pc (see Supplementary Section 9). By screening a variety of defect sites and configurations, we identify an edge-on Pc arrangement at $\mathrm{V_B} - \mathrm{V_N}$ divacancies as one of the most energetically favorable Pc–defect complexes (Fig.2a). In this configuration, the spin density is confined toward the hBN plane, increasing $D$, while the concomitant reduction of molecular symmetry enhances the transverse parameter $E$.
Most of the surface defect candidates we considered can promote the upright orientation of Pc molecules, which is particularly efficient if the defects incorporate boron vacancies (see Supplementary Section 9 for a detailed discussion). This is consistent with previous experimental reports of edge-on Pc on defective hBN \cite{Guender2020}. 

To determine the spin configuration of Pc, we performed vector-field mapping of the spin transitions. By applying magnetic fields along different orientations, the molecular axis can be identified through a fitting of the field-dependent ODMR response. In Fig.2b, we present the ODMR spectrum as a function of the magnetic field applied along the z-direction, i.e., the direction perpendicular to the hBN layer. The $\ket{\mathrm{T_x}}\leftrightarrow\ket{\mathrm{T_y}}$ and $\ket{\mathrm{T_y}}\leftrightarrow\ket{\mathrm{T_z}}$ transitions shift in opposite directions. By combining the responses for the other two orthogonal field orientations (See Supplementary Section 6), we are able to unequivocally demonstrate that the observed ODMR spots originate from Pc molecules with edge-on orientations on the hBN flakes as schematically shown in Fig.1a. Raman spectra recorded at these ODMR-active spots were compared with Raman spectra acquired on the bare substrate area (Supplementary Section 5). The mode not observed on the bare substrate but near the ODMR active spot at 1533~cm$^{-1}$ is a clear Raman signature of the edge-on configuration \cite{Seto2012}, corroborating the previous ODMR assignment. Here, the ODMR contrast remains clearly resolved for the intermediate spin-triplet manifold even under a large magnetic field (see Supplementary Section 6) applied off the molecular quantization axis (molecular z axis). This enables vector-resolved magnetic sensing with a significantly expanded dynamic range, overcoming the limitations of standard NV magnetometry where ODMR contrast diminishes under large off-axis magnetic fields.

%We now turn our attention to the role of hBN surface defects. Samples with three different surface treatments were studied in order to identify a correlation between the surface defect density and the occurrence of ODMR spots: pristine hBN with no treatment, electron irradiated hBN samples, and O$_2$-plasma treated samples. The hBN samples with heavy O$_2$-plasma bombardment show a strong increase in the number of ODMR spots, whereas in electron-irradiated samples, only a few spots are observed. In pristine high-quality hBN, almost none can be detected (See supplementary section 2 for more details). 

To understand how Pc interacts with hBN, we first determine its molecular orientation using a polarization-resolved measurement. Because the Pc optical transition dipole lies along the molecular y-axis, as illustrated in Fig.2a, its absorption is intrinsically linearly polarized in this direction \cite{Hinderhofer2007}. Using this property, we control the excitation polarization by rotating a half-wave plate. By measuring the resulting polarization-dependent fluorescence, we can reliably determine the molecular axis of Pc on hBN. Statistical data collected on over 100 spin-active sites reveal that the orientation of the molecules is clustered into three groups (Fig.2d), one aligned with the long boundary of the hBN flake, referred to as 0$^\circ$, and the other near 60$^\circ$ and 120$^\circ$, reflecting the threefold symmetry of the hBN lattice. These data are shown in Fig.2c together with three polarization maps representative of each group. In view of the three-fold symmetry and since exfoliated hBN flakes frequently exhibit straight edges along one of their crystalline directions, the y-axis of the ODMR active molecules apparently aligns with the hBN bonds. This corroborates the scenario that such molecules interact strongly with the hBN, possibly via two adjacent defect sites $\mathrm{V_B} - \mathrm{V_N}$ (although alternative configurations are also conceivable; see Supplementary Section 9), where this interaction drives self-oriented molecular assembly.

\section*{Lifetime-Limited Coherence Towards Surface Sensing}
Since the Pc stands upright on the hBN surface, it is spatially lifted from the plane of boron and nitrogen atoms. This effectively decouples from the nuclear spins of the hBN substrate, and the coherence is limited mainly by the protons in the molecule itself. Hence, the Pc experiences a much cleaner magnetic environment compared to an intrinsic defect in hBN \cite{Haykal2022}, as well as Pc embedded in an organic host, where additional nuclear spins from the matrix contribute to the decoherence. This simplifies the system considerably, since other nuclear sources only contribute marginally. It also turns this configuration particularly suitable for investigating whether isotope engineering and dynamical decoupling (DD) sequences can mitigate remaining sources of decoherence to further enhance performance.

Since the gyromagnetic ratio of deuterium is 6.5 times smaller than that of hydrogen, we considered deuterated Pc (Pc-D$_{14}$) to benefit from the reduced hyperfine coupling as illustrated in Fig.3a, and indeed, at 4~K, a coherence time $T_2$ of $39.8~\upmu$s is observed for the $\ket{\mathrm{T_y}} – \ket{\mathrm{T_z}}$ manifold (Fig.3b). This is a more than 10-fold improvement over the $T_2$ for Pc-H$_{14}$. The oscillations in the Pc-D$_{14}$ Hahn-echo traces likely arise from quadrupolar coupling of deuterium or other nuclei associated with the substrate. We also implemented dynamical decoupling to suppress the impact of low frequency noise using the Carr–Purcell–Meiboom–Gill (CPMG) sequences. A systematic increase is observed when dynamical decoupling pulses are applied. In our spin-protection scheme shown in Fig.3c, the $\ket{\mathrm{T_x}}$ and $\ket{\mathrm{T_z}}$ states are long-lived triplet sublevels for spin manipulation, whereas the $\ket{\mathrm{T_y}}$ state is short-lived and is used to perform the readout (See detailed characterizations of lifetime in Supplementary Section 7). In deuterated Pc, we have achieved a $T_{2,\mathrm{DD}}$ exceeding $300~\upmu$s with this configuration (Fig.3d). Even more, both $\ket{\mathrm{T_y}} \leftrightarrow \ket{\mathrm{T_z}}$ and $\ket{\mathrm{T_x}} \leftrightarrow \ket{\mathrm{T_z}}$ transitions can reach the lifetime limitation of each spin manifold as summarized in Fig.3e (See detailed characterization in Supplementary Section 7). Dynamical decoupling also prolongs the coherence of normal Pc (H$_{14}$) to $T_{2,\mathrm{DD}} \approx 130~\upmu$s as shown in Supplementary Section 7. These values represent record-long spin coherence times for hBN-related systems and are obtained despite being at the surface. These values are comparable to, or even surpass, those for implanted spin defects, generated in bulk crystals including diamond and SiC \cite{Naydenov2010,Kasper2020,Lin2021}.

The robustness of their spin properties enables the Pc molecules to function as quantum sensors for detecting their surroundings. Here, we demonstrate that it's possible to use Pc-H$_{14}$ to detect their own proton nuclear spins in a Hahn echo experiment at room temperature. In Fig.4a, we plot the time evolution of the ODMR photoluminescence signal for six different magnetic fields, reflecting degrees of electron spin polarization. The hyperfine interaction causes a periodic revival of the spin polarization with the periodicity determined by the Larmor frequency of the nuclei, and the decay unveils the coherence time. Fits to these Hahn-echo decays yield a gyromagnetic ratio of $\sim$43~MHz/T matching that of the protons in Pc (Fig.4b).

Finally, as proof of principle that these molecular sensors on hBN can be integrated into a device architecture for sensing 2D magnetic layer materials, we fabricated an hBN-encapsulated \ce{Fe3GaTe2} (FGT) van der Waals heterostructure and deposited Pc molecules on the top hBN surface, as illustrated in Fig.4c. FGT has been recently identified as a room-temperature 2D ferromagnet \cite{Zhang2022}. The hBN spacer defines the sensor–sample distance, allowing the molecular spins to detect magnetic responses from the underlying magnetic layer under ambient conditions (see more characterizations in Supplementary Section 8). Figure 4d illustrates that molecules located near the FGT region exhibit clear ODMR shifts compared with those sitting on the bare substrate. The observed shift reflects a local magnetic response of roughly 1~mT. In this hybrid configuration, the ensemble of molecular spins operates as an integrated quantum sensor that uncovers the underlying magnetic behavior of the 2D magnet.

\section*{Conclusions}
In conclusion, we have demonstrated the operation of an optically addressable spin sensor directly at the interface by combining pentacene molecules with the 2D material hBN. The spin lifetime, spin coherence time, as well as the optical properties required for sensing, are preserved from 4~K up to room temperature. The spin coherence time not only surpasses that of molecular spin sensors embedded in an organic bulk matrix, but also exceeds those of all previously reported spin defects in hBN by more than an order of magnitude \cite{Gottscholl2020,Mendelson2020,Chejanovsky2021,Stern2022,Gao2025}. Despite its position at the surface, this spin sensor reaches a coherence level comparable to that of many well-established spin defects that need to be buried inside the host material to avoid performance degradation. Particularly noteworthy is the excellent photostability of Pc spins at ODMR-active sites. The signals persist for days in ambient conditions under continuous laser illumination. This is not the case for organic molecules in a thin-film bulk matrix or solutions that suffer from severe photobleaching \cite{Widengren1996,Demchenko2020}. Apparently, the hBN host plays a stabilizing role, which may inspire new strategies for realizing other robust molecular sensors.

These hybrid sensors are inherently compatible with 2D-material fabrication and surface-NMR platforms, so that a seamless integration with a diverse set of device architectures is conceivable. By using wafer-scale hBN as a substrate, our molecular quantum system can be scaled to several inches with deterministic positioned molecules, opening new opportunities for quantum sensing. The device geometry, where the hBN covered with molecules is placed on top of the material to be studied, may benefit from a reduced hBN layer thickness in order to enhance the coupling strength of the spin sensor beyond the dipolar limit and enter the exchange-interaction regime \cite{Kovarik2024}. Pc represents only one example among many aromatic molecules that possess intermediate triplet states. A vast family of more than 10,000 related molecules exists with tunable chemical modifications \cite{Mann2025,Anthony2001,Yamada2005}, and hence the surface spin sensor concept put forward here can be extended to numerous other candidate molecules.

\clearpage
\noindent
\section*{Acknowledgments}
We acknowledge funding support from the European Union through the project C-QuEnS (Grant No. 101135359); the BMBF through the projects QCOMP (Grant No. 03ZU1110HA), QSOLID (Grant No. 13N16159), and QMAT (Grant No. 03ZU2110HA); and the Zeiss Foundation through QPhoton. J.H.S. acknowledges support from the DFG Priority Program SPP 2244. A.G., A.P., G.B., and R.M. acknowledge support from the Quantum Information National Laboratory of Hungary, funded by the National Research, Development, and Innovation Office (NKFIH) under Grant No. 2022-2.1.1-NL-2022-00004. A.G. further acknowledges access to high-performance computational resources provided by KIFÜ (Governmental Agency for IT Development, Hungary), and funding from the European Commission through the QuSPARC (Grant No. 101186889) and SPINUS (Grant No. 101135699) projects. A.P. acknowledges the financial support of a János Bolyai Research Fellowship of the Hungarian Academy of Sciences. R.M. is grateful for support from the Stipendium Hungaricum scholarship.

\section*{Author Contributions}
R.P. and J.W. conceived the project. R.P., X.Z., and Y.K. fabricated Pc-hBN devices. X.Z., Y.K., C.C., and R.P. performed room-temperature characterization with assistance from K.W., Y.S., and C.H. X.Z., Y.K., C.C., and K.W. carried out cryogenic measurements, assisted by V.B., M.L., and S.J. C.Y. performed all AFM scans. N.G. and T.P. performed the Raman measurements. G.B., R.M., A.P., and A.G. performed the DFT analysis of molecule interactions. R.P. and J.H.S. wrote the paper, assisted by J.W., X.Z., Y.K., C.C., N.G., G.B., and A.G. R.P. supervised the research. All authors discussed the results and commented on the paper.

\clearpage
\bibliography{scibib}

\bibliographystyle{science}

% Following is a new environment, {scilastnote}, that's defined in the
% preamble and that allows authors to add a reference at the end of the
% list that's not signaled in the text; such references are used in
% *Science* for acknowledgments of funding, help, etc.

% For your review copy (i.e., the file you initially send in for
% evaluation), you can use the {figure} environment and the
% \includegraphics command to stream your figures into the text, placing
% all figures at the end.  For the final, revised manuscript for
% acceptance and production, however, PostScript or other graphics
% should not be streamed into your compliled file.  Instead, set
% captions as simple paragraphs (with a \noindent tag), setting them
% off from the rest of the text with a \clearpage as shown  below, and
% submit figures as separate files according to the Art Department's
% instructions.

\clearpage

\begin{figure}[h!]
  \centering
  \includegraphics[width=1\textwidth]{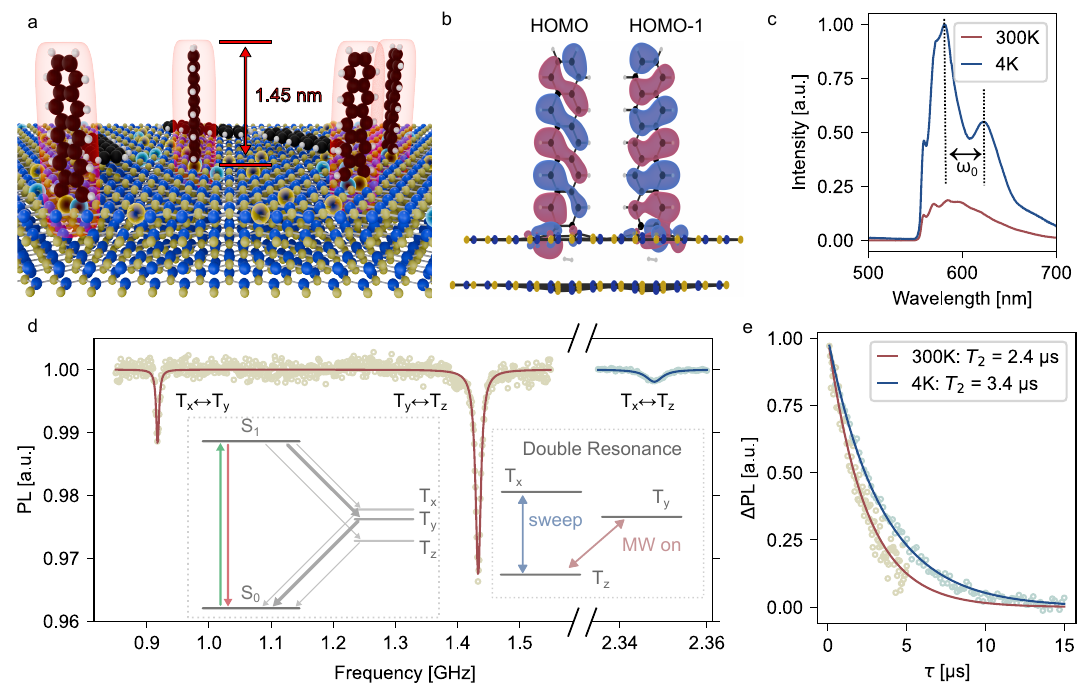}
  \caption{\textbf{Surface molecular spins hosted on hBN.}
\textbf{a}, Schematic illustration of Pc molecules coupled to hBN surface defects in the edge-on configuration, which enables optical addressing of spins. 
\textbf{b}, Spatial distributions of the highest occupied molecular orbitals (HOMO) of the triplet spins.
\textbf{c}, Photoluminescence spectra of Pc molecules on hBN at 4 K and room temperature. $\omega_0$ denotes a characteristic molecular phonon mode. 
\textbf{d}, ODMR spectrum of Pc at ambient conditions, revealing the $\ket{\mathrm{T_x}} – \ket{\mathrm{T_y}}$ and $\ket{\mathrm{T_y}} – \ket{\mathrm{T_z}}$ transitions at 917~MHz and 1433~MHz, respectively. The $\ket{\mathrm{T_x}} – \ket{\mathrm{T_z}}$ transition is accessed via a double-resonance scheme, as illustrated in the right inset, where two microwave tones are applied.
\textbf{e}, Hahn-echo spin-coherence measurements of Pc spins at room temperature and 4~K, yielding a characteristic $T_2$ time of 3.4~$\upmu$s at 4K.
}
  \label{fig:figure1}
\end{figure}

\clearpage
% The probability of [-30,29], [30,89], [90,149] is 0.3408, 0.4627, 0.1965

\begin{figure}[h!]
  \centering
  \includegraphics[width=1\textwidth]{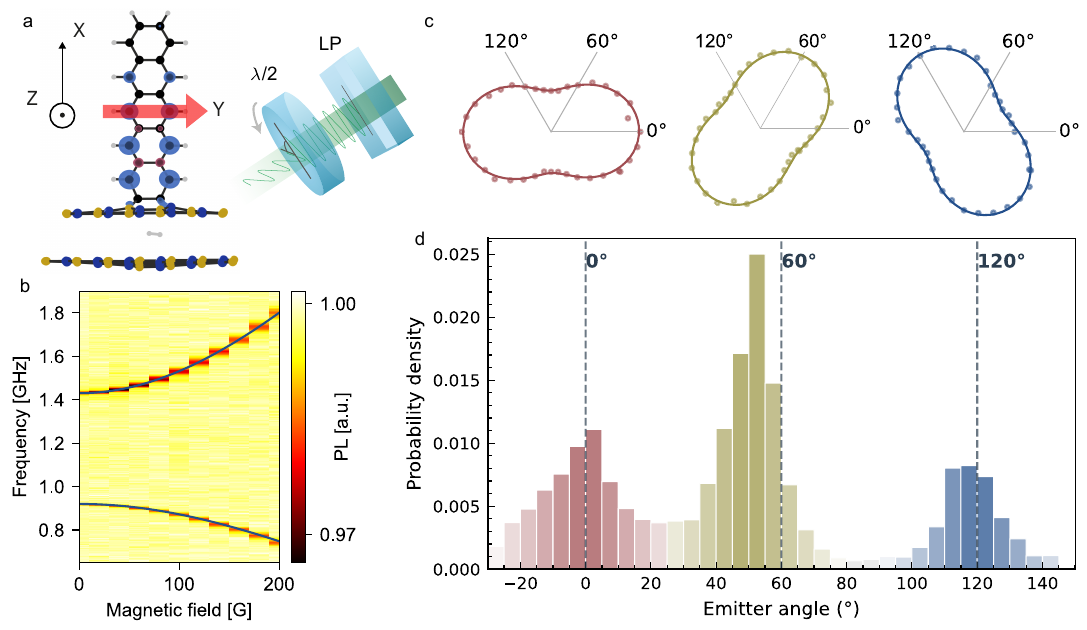}
  \caption{\textbf{Pc configuration on hBN surfaces.}
\textbf{a}, Calculated spin-density distribution of Pc, showing a shift of spin density toward the hBN plane, which leads to an increased zero-field splitting with optical dipole orientation of the Pc aligned along the molecular Y axis. The molecular axis orientation is indicated according to the conventional definition. Right: schematic of the polarization-resolved measurement used to determine the Pc molecular Y-axis orientation.
%Representative confocal image of an hBN surface fully covered with Pc molecules. Every fluorescent spot exhibits a clear ODMR spectrum and linear dipole polarization. The white line marks the edge of the hBN flake; scale bar, 5 $\upmu$m.
\textbf{b}, Out-of-plane Magnetic-field dependence of the $\ket{\mathrm{T_x}} – \ket{\mathrm{T_y}}$ and $\ket{\mathrm{T_y}} – \ket{\mathrm{T_z}}$ transitions, consistent with an edge-on configuration of Pc molecules. 
\textbf{c}, Representative polarization maps for Pc orientations at $0^{\circ}$, $60^{\circ}$, and $120^{\circ}$ with respect to the hBN edges show clear linear polarization behavior.
\textbf{d}, Histogram of hundreds of Pc ODMR spots, revealing the threefold symmetry of Pc alignment relative to the underlying hBN lattice. 
}
  \label{fig:figure2}
\end{figure}

\clearpage
\begin{figure}[h!]
  \centering
  \includegraphics[width=1\textwidth]{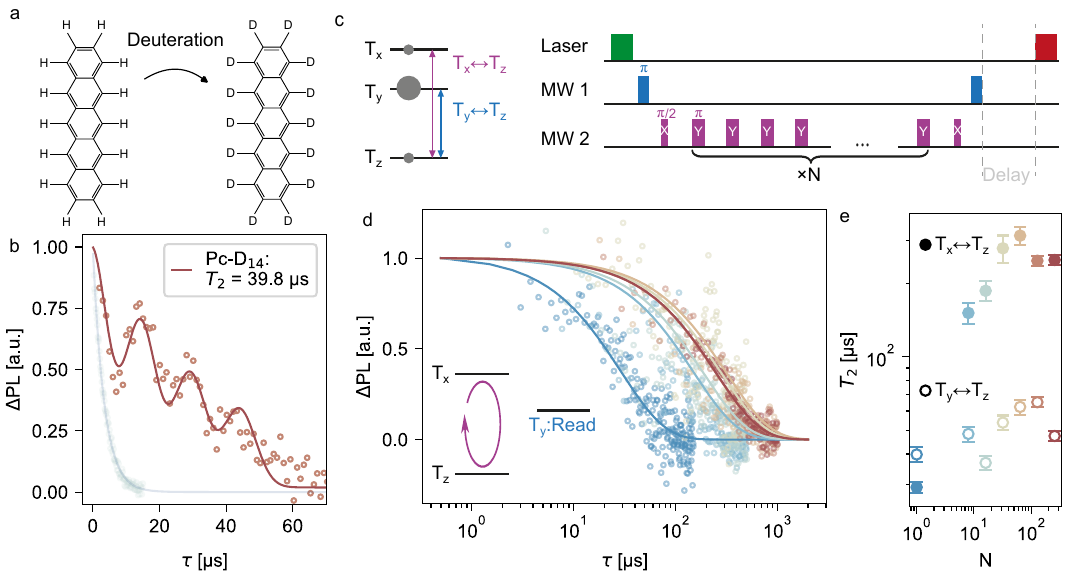}
  \caption{\textbf{Engineering of spin coherence toward the lifetime limit.}
\textbf{a}, Schematic illustration of molecular deuteration from Pc-H$_{14}$ to Pc-D$_{14}$, which suppresses nuclear-spin-induced decoherence. 
\textbf{b}, The Hahn-echo spin coherence time ($T_2$) in Pc-D$_{14}$ exhibits a tenfold enhancement relative to Pc-H$_{14}$. For comparison, the Pc-H$_{14}$ data are plotted as a semi-transparent reference curve.
\textbf{c}, Schematic of dynamical decoupling pulse sequences (CPMG) used to protect spin coherence in the molecular $\ket{\mathrm{T_x}} – \ket{\mathrm{T_z}}$ basis with $\mathrm{T_y}$ for readout. 
\textbf{d}, Extended spin coherence of Pc-D$_{14}$ under dynamical decoupling, with spins protected in the $\ket{\mathrm{T_x}} – \ket{\mathrm{T_z}}$ manifold. 
\textbf{e}, The prolongation of the coherence time as a function of the number of applied $\pi$ pulses N. For both the $\ket{\mathrm{T_y}} – \ket{\mathrm{T_z}}$ and $\ket{\mathrm{T_x}} – \ket{\mathrm{T_z}}$ transitions, the spin coherence approaches the lifetime limit for their own spin manifolds.
}
  \label{fig:figure3}
\end{figure}

\clearpage
\begin{figure}[h!]
  \centering
  \includegraphics[width=1\textwidth]{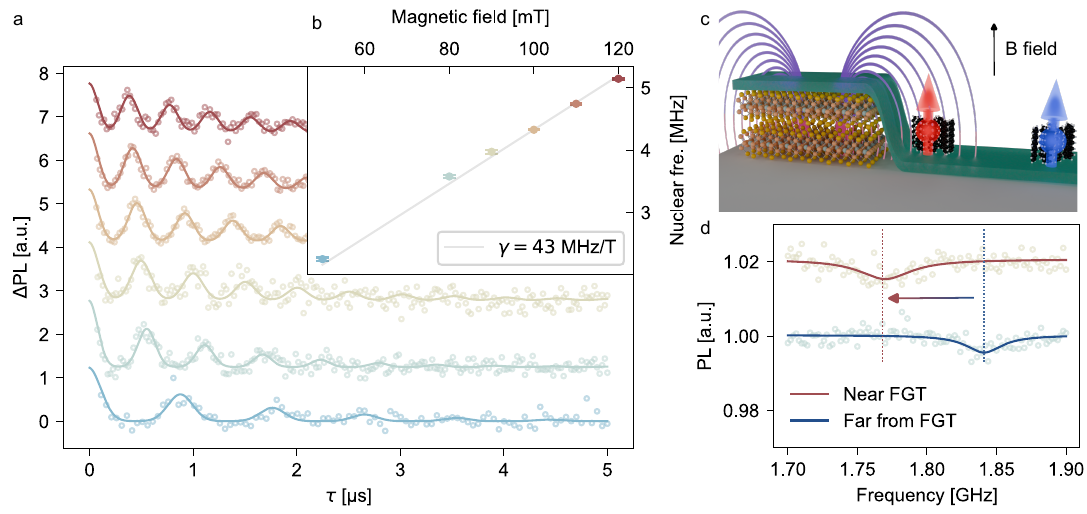}
  \caption{\textbf{Exploiting Pc as a room temperature spin surface sensor.}
\textbf{a}, Hahn-echo measurements recorded for six different values of the magnetic field in order to detect Pc’s own proton spins. $\Delta$PL reflects the degree of spin polarization. The observed revivals in spin polarization indicate coupling to nearby nuclear spins. 
\textbf{b}, Field dependence of the nuclear precession frequency, showing an effective gyromagnetic ratio close to that of the protons. 
\textbf{c}, Illustration of the two-dimensional (2D) sensing architecture, where Pc molecules are deposited on an hBN layer
that itself covers a 2D ferromagnet (Fe$_3$GaTe$_2$). 
\textbf{d}, ODMR spectrum of Pc molecules located in the hBN area covering the  2D magnet. Also shown is a reference spectrum on hBN far away from FGT. The arrow marks the observed shift induced by the magnetic layer.
}
  \label{fig:figure4}
\end{figure}
\end{document}